# Topological Gauging of $N = 16$ Supergravity in Three-Dimensions

Hitoshi NISHINO[1] and Subhash RAJPOOT[2]

*Department of Physics & Astronomy*
*California State University*
*Long Beach, CA 90840*

### Abstract

We present a topologically non-trivial generalization of gauged $N = 16$ supergravity on the coset $E_{8(+8)}/SO(16)$ in three-dimensions. This formulation is based on a combination of $BF$-term and a Chern-Simons term for an $SO(16)$ gauge field $A_\mu{}^{IJ}$. The fact that an additional vector field $B_\mu{}^{IJ}$ is physical and propagating with couplings to $\sigma$-model fields makes our new gauging non-trivial and different from the conventional one. Even though the field strength of the $A_\mu{}^{IJ}$-field vanishes on-shell, the action is topologically non-trivial due to non-vanishing $\pi_3$-homotopy. We also present an additional modifications by an extra Chern-Simons term. As by-products, we give also an application to $N = 9$ supergravity coupled to a $\sigma$-model on the coset $F_{4(-20)}/SO(9)$, and a new $BF$-Chern-Simons theory coupled to $^\forall N$ extended supergravity.

---

[1]E-Mail: hnishino@csulb.edu
[2]E-Mail: rajpoot@csulb.edu

## 1. Introduction

Recently, there has been a considerable development for $N = 16$ maximally gauged supergravity in three-dimensions (3D) on the coset $E_{8(+8)}/SO(16)$ [1][2]. This is due to the unexpectedly rich structure of the gauged theory on such a huge coset $E_{8(+8)}/SO(16)$, which can not be easily explained by conventional dimensional reductions [3] from 11D supergravity [4]. In fact, new generalizations of simple dimensional reduction by Scherk-Schwarz [3] have been recently discovered [5], leading to various mass parameters, presumably corresponding to distinct gauging schemes in 3D [2]. Such relationships with 11D by dimensional reductions provide one important motivation of the study of $N = 16$ supergravity in 3D, since 11D supergravity is regarded as the low energy limit of M-theory [6]. An additional motivation for the study of $N = 16$ supergravity in 3D is the link between 1D supersymmetric matrix model [7] and M-theory, because the former may be further promoted to supersymmetric 3D model [8][9].

As suggested in [2], another important aspect is that the rich structures of gauged $N = 16$ supergravity in 3D indicate the existence of supergravity theories even in dimensions higher than 11D, such as F-theory in 12D [10] or S-theory in 13D [11]. From these viewpoints, it is natural to expect the existence of some higher-dimensional supergravity even beyond 11D [12][13], which can be studied by investigating gauged $N = 16$ supergravity in 3D [2], or its other possible generalizations.

On the other hand, there has been an independent development related to supersymmetric Chern-Simons theories in 3D [14], in which it has been shown that certain Chern-Simons theories in 3D can exist for arbitrary number of supersymmetries up to infinity [15]. Even though these theories are constructed in the absence of $\sigma$-models on such a coset as $E_{8(+8)}/SO(16)$, these results indicate that there may still exist many other overlooked non-trivial gauge theories in 3D, even for maximally extended supergravity in the presence of $\sigma$-models. The importance of Chern-Simons theory in 3D has been stressed also in different contexts, such as the conjectural relationship between $SU(N)$ Chern-Simons theory on $S^3$ and topological string with a non-compact Calabi-Yau threefold as the target space [16].

Considering these developments in extended supergravity in 3D, it seems important to address a question whether there is any other generalization of gauging maximal[3] supergravity in 3D, related to possible Chern-Simons terms. In this paper we show such an alternative way of gauging of $N = 16$ supergravity in 3D. We introduce the minimal gauge coupling to an independent $SO(16)$ gauge field $A_\mu$, distinct from $B_\mu{}^{IJ}$ used in the conventional gauging [2], together with a $BF$-term and a Chern-Simons term in the lagrangian. Even though the former forces the field strength of $A_\mu$ to vanish, due to the non-trivial feature of the Chern-Simons term in 3D with $\pi_3(SO(16)) = \mathbb{Z}$, the system has topologically non-trivial configurations. We also show that we can add an independent $SO(16)$ gauge field that can form an additional non-trivial Chern-Simons term. As an application, we show a similar mechanism in the case of $N = 9$ supergravity with the coset $F_{4(-20)}/SO(9)$.

---
[3]The word 'maximal' here means the maximal $N = 16$ supergravity in the presence of $\sigma$-model. If there is no $\sigma$-model, there is no limit for $N$, as indicated in [14].

This paper is organized as follows: We start with the topological gauging as described, with the two new terms of $BF$-type and a Chern-Simons type in the next section. In section 3, we apply a similar technique to the case of $N = 9$ supergravity coupled to an $F_{4(-20)}/SO(9)$ $\sigma$-model. As a by-product, we also give a new supersymmetric $SO(N)$ Chern-Simons lagrangian coupled to $\forall N$ extended supergravity in the absence of $\sigma$-model, that was not given before. Concluding remarks will be given in section 5. Appendix A is devoted to $\Gamma$-matrix properties for $SO(9)$ Clifford algebra, while Appendix B is for Fierz identities for $SO(9)$ Majorana spinors.

## 2. Topological $SO(16)$-Gauging

We now consider a new gauging of $SO(16)$ in $N = 16$ supergravity in 3D coupled to $E_{8(+8)}/SO(16)$ $\sigma$-model. Our formulation of $SO(16)$-gauging is a generalization of the non-gauged theory [1] by two important new terms: One $BF$-term and one Chern-Simons term. We introduce two different vector fields $A_\mu{}^{IJ}$ and $B_\mu{}^{IJ}$, but only the former is the gauge field of $SO(16)$, coupling minimally to the gravitino, while the latter appears only in the $BF$-term. We also introduce an extra gauge field $C_\mu{}^{IJ}$ of $SO(16)$, with an extra Chern-Simons term.

Our field content is $(e_\mu{}^m, \psi_\mu{}^I, \chi_{\dot{A}}, \varphi_A, A_\mu{}^{IJ}, B_\mu{}^{IJ}, C_\mu{}^{IJ}, \lambda^I)$. The first four fields are the same as those in the non-gauged case in [1][2], while our new fields are $A_\mu{}^{IJ}$, $C_\mu{}^{IJ}$ and $\lambda^I$. The field $B_\mu{}^{IJ}$ has a similar supersymmetry transformation rule to that in [2], but its couplings to other fields are different, as will be seen. Our total lagrangian is[4]

$$e^{-1}\mathcal{L}_{16} = -\tfrac{1}{4}R + \tfrac{1}{2}e^{-1}\epsilon^{\mu\nu\rho}(\overline{\psi}_\mu{}^I \mathcal{D}_\nu(\omega,\mathcal{Q})\psi_\rho{}^I) + \tfrac{1}{4}g^{\mu\nu}\mathcal{P}_{\mu A}\mathcal{P}_{\nu A}$$
$$- \tfrac{i}{2}(\overline{\chi}_{\dot{A}} \gamma^\mu \mathcal{D}_\mu(\omega,\mathcal{Q})\chi_{\dot{A}}) - \tfrac{1}{2}(\Gamma^I)_{A\dot{B}}(\overline{\psi}_\mu{}^I \gamma^\nu \gamma^\mu \chi_{\dot{B}})\mathcal{P}_{\nu A}$$

---

[4] Our notation is such as $(\eta_{mn}) =$ diag. $(-,+,+)$, $\epsilon^{012} = +1$, $\gamma^{mnr} = -i\epsilon^{mnr}$, $\gamma^{mn} = -i\epsilon^{mnr}\gamma_r$, $2\gamma^m = +i\epsilon^{mnr}\gamma_{nr}$. The $m, n, \cdots = (0), (1), (2)$ are the local Lorentz, while $\mu, \nu, \cdots = 0, 1, 2$ are curved indices. These are essentially the same as [2], except that we always use subscripts for the spinorial indices $A, B, \cdots$ for the **128**, or the dotted ones $\dot{A}, \dot{B}, \cdots$ for the $\overline{\mathbf{128}}$ of $SO(16)$, and upper case for the **16**-indices $I, J, \cdots$. Note also that our gauged group is $SO(16)$ instead of arbitrary subgroup of $E_{8(+8)}$ as in [2].

$$+ \tfrac{1}{2} g e^{-1} \epsilon^{\mu\nu\rho} B_\mu{}^{IJ} F_{\nu\rho}{}^{IJ} + \tfrac{1}{2} m e^{-1} \epsilon^{\mu\nu\rho} (F_{\mu\nu}{}^{IJ} A_\rho{}^{IJ} + \tfrac{2}{3} g A_\mu{}^{IJ} A_\nu{}^{JK} A_\rho{}^{KI})$$

$$+ \tfrac{1}{2} \widetilde{m} e^{-1} \epsilon^{\mu\nu\rho} (H_{\mu\nu}{}^{IJ} C_\rho{}^{IJ} + \tfrac{2}{3} g C_\mu{}^{IJ} C_\nu{}^{JK} C_\rho{}^{KI}) + \tfrac{1}{2} \widetilde{m} (\overline{\lambda}^I \lambda^I)$$

$$- \tfrac{1}{8} (\overline{\psi}_\rho{}^I \gamma^{\rho\sigma\tau} \psi_\sigma{}^J)(\overline{\chi} \gamma_\tau \Gamma^{IJ} \chi) + \tfrac{1}{8} (\overline{\psi}_\mu{}^I \gamma_\nu \psi^{\mu J})(\overline{\chi} \gamma_\nu \Gamma^{IJ} \chi)$$

$$- \tfrac{1}{8} (\overline{\psi}_\mu{}^I \gamma^\nu \gamma^\mu \psi_\nu{}^J)(\overline{\chi}\chi) + \tfrac{1}{8} (\overline{\chi}\chi)^2 - \tfrac{1}{96} (\overline{\chi} \gamma_\mu \Gamma^{IJ} \chi)^2 \quad . \tag{2.1}$$

Our action $I_{16} \equiv \int d^3 x \, \mathcal{L}_{16}$ is invariant under supersymmetry

$$\delta_Q e_\mu{}^m = + i(\overline{\epsilon}^I \gamma^m \psi_\mu{}^I) \quad , \tag{2.2a}$$

$$\delta_Q \psi_\mu{}^I = + \mathcal{D}_\mu(\widehat{\omega}, \mathcal{Q})\epsilon^I - \tfrac{i}{4}(\gamma^\nu \epsilon^J)(\overline{\chi} \Gamma^{IJ} \gamma_{\mu\nu} \chi) - \Sigma^{IJ} \psi_\mu{}^J$$

$$+ g e^{-1} \epsilon_\mu{}^{\rho\sigma} \epsilon^J \widehat{G}_{\rho\sigma}{}^{IJ} + i g \gamma^\nu \epsilon^J \widehat{G}_{\mu\nu}{}^{IJ}$$

$$+ \tfrac{i}{2} g e^{-1} \epsilon_\mu{}^{\rho\sigma} \gamma_\rho \epsilon^J \widehat{\mathcal{P}}_{\sigma A} \mathcal{V}_A{}^{IJ} + g \epsilon^J \widehat{\mathcal{P}}_{\mu A} \mathcal{V}_A{}^{IJ}$$

$$+ \tfrac{1}{2} g e^{-1} \epsilon_\mu{}^{\rho\sigma} \gamma_\rho \epsilon^J (\overline{\chi} \gamma_\sigma \Gamma^{IJ} \chi) - \tfrac{i}{4} g \epsilon^J (\overline{\chi} \Gamma^{IJ} \gamma_\mu \chi)$$

$$+ 2 i m (\gamma^\nu \epsilon^J) F_{\mu\nu}{}^{IJ} + 2 m e^{-1} \epsilon_\mu{}^{\nu\rho} \epsilon^J F_{\nu\rho}{}^{IJ} \equiv \widehat{\mathcal{D}}_\mu \epsilon^I \quad , \tag{2.2b}$$

$$\delta_Q \chi_{\dot{A}} = + \tfrac{i}{2} (\Gamma^I)_{B\dot{A}} (\gamma^\mu \epsilon^I) \widehat{\mathcal{P}}_{\mu B} - \tfrac{1}{4} (\Gamma^{IJ} \chi)_{\dot{A}} \Sigma^{IJ} \quad , \tag{2.2c}$$

$$\delta_Q \varphi = + \frac{\Phi}{\sinh \Phi} (\overline{\epsilon}^I \Gamma^I \chi) \equiv \left( \frac{\Phi}{\sinh \Phi} \right) S \quad , \quad \tfrac{1}{2} \Sigma^{IJ} X^{IJ} \equiv \left( \tanh \frac{\Phi}{2} \right) S \quad , \tag{2.2d}$$

$$\delta_Q A_\mu{}^{IJ} = + i(\overline{\epsilon}^{[I|} \gamma^\nu \widehat{\mathcal{R}}_{\mu\nu}{}^{|J]}) + e^{-1} \epsilon_\mu{}^{\rho\sigma} (\overline{\epsilon}^{[I} \widehat{\mathcal{R}}_{\rho\sigma}{}^{J]}) + i e^{-1} \epsilon_\mu{}^{\rho\sigma} (\overline{\epsilon}^{[I} \Gamma^{J]} \gamma_\rho \chi) \widehat{\mathcal{P}}_{\sigma A} \quad , \tag{2.2e}$$

$$\delta_Q B_\mu{}^{IJ} = + \tfrac{1}{2} (\overline{\epsilon}^K \psi_\mu{}^L) \mathcal{V}^{KL,IJ} - \tfrac{i}{2} (\overline{\epsilon}^K \Gamma^K \gamma_\mu \chi)_A \mathcal{V}_A{}^{IJ} \quad , \tag{2.2f}$$

$$\delta_Q C_\mu{}^{IJ} = + i(\overline{\epsilon}^{[I} \gamma_\mu \lambda^{J]}) \quad , \tag{2.2g}$$

$$\delta_Q \lambda^I = + \gamma^{\mu\nu} \epsilon^J H_{\mu\nu}{}^{IJ} - \tfrac{i}{2} (\overline{\epsilon}^J \gamma^\mu \psi_\mu{}^J) \lambda^I \quad , \tag{2.2h}$$

where we use the operator symbol $\Phi$ defined by $\Phi \alpha \equiv [\varphi, \alpha]$ for an arbitrary $E_8$ generator-valued field $\alpha$ with $\varphi \equiv \varphi_A Y_A$ for the generators $Y_A$ on the coset $E_{8(+8)}/SO(16)$ [1][2]. Our covariant derivatives are such as

$$\mathcal{D}_{[\mu}(\widehat{\omega}, \mathcal{Q}) \psi_{\nu]}{}^I \equiv D_{[\mu}(\widehat{\omega}) \psi_{\nu]}{}^I + \mathcal{Q}_{[\mu}{}^{IJ} \psi_{\nu]}{}^J \quad ,$$

$$\mathcal{V}^{-1} \mathcal{D}_\mu \mathcal{V} \equiv \mathcal{V}^{-1} \partial_\mu \mathcal{V} + \tfrac{1}{2} g A_\mu{}^{IJ} \mathcal{V}^{-1} X^{IJ} \mathcal{V} \equiv \mathcal{P}_{\mu A} Y_A + \tfrac{1}{2} \mathcal{Q}_\mu{}^{IJ} X^{IJ} \quad , \tag{2.3}$$

with the generators $X^{IJ}$ of $SO(16)$ and $Y_A$ on the coset $E_{8(+8)}/SO(16)$, satisfying

$$[X^{IJ}, X^{KL}] = 2 \delta^{I[K} X^{L]J} - 2 \delta^{J[K} X^{L]I} \quad ,$$

$$[X^{IJ}, Y_A] = -\tfrac{1}{2} (\Gamma^{IJ})_{AB} Y_B \quad , \quad [Y_A, Y_B] = \tfrac{1}{4} (\Gamma^{IJ})_{AB} X^{IJ} \quad . \tag{2.4}$$

Accordingly, the gravitino field strength is

$$\mathcal{R}_{\mu\nu}{}^I \equiv \mathcal{D}_\mu(\widehat{\omega}, \mathcal{Q}) \psi_\nu{}^I - \mathcal{D}_\nu(\widehat{\omega}, \mathcal{Q}) \psi_\mu{}^I \quad . \tag{2.5}$$

Other field strengths are defined by

$$F_{\mu\nu}{}^{IJ} \equiv 2\partial_{[\mu}A_{\nu]}{}^{IJ} + 2gA_{[\mu}{}^{[I|K}A_{\nu]}{}^{K|J]} ,$$
$$G_{\mu\nu}{}^{IJ} \equiv 2\partial_{[\mu}B_{\nu]}{}^{IJ} + 4gA_{[\mu}{}^{[I|K}B_{\nu]}{}^{K|J]} ,$$
$$H_{\mu\nu}{}^{IJ} \equiv \partial_{[\mu}C_{\nu]}{}^{IJ} + 2gC_{[\mu}{}^{IK}C_{\nu]}{}^{KJ} . \quad (2.6)$$

As usual in supergravity [17], we use the 'hat' symbols for supercovariantized field strengths

$$\widehat{\mathcal{R}}_{\mu\nu}{}^{I} \equiv \widehat{\mathcal{D}}_{\mu}\psi_{\nu}{}^{I} - \widehat{\mathcal{D}}_{\nu}\psi_{\mu}{}^{I} ,$$
$$\widehat{\mathcal{P}}_{\mu A} \equiv \mathcal{P}_{\mu A} - (\Gamma^{I})_{A\dot{B}}(\overline{\psi}_{\mu}{}^{I}\chi_{\dot{B}}) \equiv \mathcal{P}_{\mu A} - (\overline{\psi}_{\mu}{}^{I}\Gamma^{I}\chi)_{A} ,$$
$$\widehat{F}_{\mu\nu}{}^{IJ} \equiv F_{\mu\nu}{}^{IJ} - 2i(\overline{\psi}_{[\mu}{}^{[I|}\gamma^{\rho}\widehat{\mathcal{R}}_{\nu]\rho}{}^{|J]}) + 2e^{-1}\epsilon_{[\mu}{}^{\rho\sigma}(\overline{\psi}_{\nu]}{}^{[I|}\widehat{\mathcal{R}}_{\rho\sigma}{}^{|J]})$$
$$\qquad + 2ie^{-1}\epsilon_{[\mu}{}^{\rho\sigma}(\overline{\psi}_{\nu]}{}^{[I}\Gamma^{J]}\gamma_{\rho}\chi)\widehat{\mathcal{P}}_{\sigma A} ,$$
$$\widehat{G}_{\mu\nu}{}^{IJ} \equiv G_{\mu\nu}{}^{IJ} - \tfrac{1}{2}(\overline{\psi}_{\mu}{}^{K}\psi_{\nu}{}^{L})\mathcal{V}^{KL,IJ} + i(\overline{\psi}_{[\mu}{}^{K}\Gamma^{K}\gamma_{\nu]}\chi)_{A}\mathcal{V}_{A}{}^{IJ} ,$$
$$\widehat{H}_{\mu\nu}{}^{IJ} \equiv H_{\mu\nu}{}^{IJ} - 2i(\overline{\psi}_{[\mu}{}^{[I}\gamma_{\nu]}\lambda^{J]}) , \quad (2.7)$$

where $\widehat{\mathcal{D}}_{\mu}$ is defined by (2.2b).

As eq. (2.6) indicates, under the local $SO(16)$ gauge symmetry, those vector fields transform as

$$\delta_{\alpha}A_{\mu}{}^{IJ} = +\partial_{\mu}\alpha^{IJ} + gA_{\mu}{}^{IK}\alpha^{KJ} + gA_{\mu}{}^{JK}\alpha^{IK} ,$$
$$\delta_{\alpha}B_{\mu}{}^{IJ} = +B_{\mu}{}^{IK}\alpha^{KJ} + B_{\mu}{}^{JK}\alpha^{IK} ,$$
$$\delta_{\alpha}C_{\mu}{}^{IJ} = +\partial_{\mu}\alpha^{IJ} + gC_{\mu}{}^{IK}\alpha^{KJ} + gC_{\mu}{}^{JK}\alpha^{IK} . \quad (2.8)$$

Even though $B_{\mu}{}^{IJ}$ is a vector, it does *not* gauge $SO(16)$, but just transforms homogeneously.

There some important geometrical relationships related to our coset $E_{8(+8)}/SO(16)$, such as the integrability conditions

$$\mathcal{Q}_{\mu\nu}{}^{IJ} + \tfrac{1}{2}(\Gamma^{IJ})_{AB}\mathcal{P}_{\mu A}\mathcal{P}_{\nu B} = gF_{\mu\nu}{}^{IJ} ,$$
$$\mathcal{D}_{\mu}\mathcal{P}_{\nu A} - \mathcal{D}_{\nu}\mathcal{P}_{\mu A} = gF_{\mu\nu}{}^{IJ}\mathcal{V}_{A}{}^{IJ} , \quad (2.9)$$

where

$$\mathcal{Q}_{\mu\nu}{}^{IJ} \equiv 2\partial_{[\mu}\mathcal{Q}_{\nu]}{}^{IJ} + 2\mathcal{Q}_{[\mu}{}^{IK}\mathcal{Q}_{\nu]}{}^{KJ} ,$$
$$\mathcal{D}_{\mu}\mathcal{P}_{\nu A} \equiv \partial_{\mu}\mathcal{P}_{\nu A} + \tfrac{1}{4}\mathcal{Q}_{\mu}{}^{IJ}(\Gamma^{IJ})_{AB}\mathcal{P}_{\nu B} , \quad (2.10)$$

Some remarks are now in order. First, the $BF$-term in our lagrangian (2.1) forces the field strength $F_{\mu\nu}{}^{IJ}$ to vanish. This is equivalent to forcing the gauge field $A_{\mu}{}^{IJ}$ to be 'pure gauge', which seems to lead to a trivial system. However, since this system is in 3D, and moreover due to the homotopy mapping $\pi_3(SO(16)) = \mathbb{Z}$, there are some topologically non-trivial configuration possible at the action level, *via* the explicit $A^3$-term in (2.1).

Second, the gauge field $C_\mu{}^{IJ}$ can be added as an extra gauge field, transforming differently from $A_\mu{}^{IJ}$, but it has no direct interactions with other fields. Its associated Chern-Simons term $HC \oplus C^3$ will be also non-trivial due to $\pi_3(SO(16)) = \mathbb{Z}$.

Third, the $A_\mu{}^{IJ}$-field equation[5]

$$e^{-1}\epsilon_\mu{}^{\rho\sigma}\widehat{G}_{\rho\sigma}{}^{IJ} \doteq -\mathcal{V}_A{}^{IJ}\widehat{\mathcal{P}}_{\mu A} + \tfrac{i}{4}(\overline{\chi}\gamma_\mu\Gamma^{IJ}\chi) \ , \tag{2.11}$$

is exactly the same as the $B_\mu{}^{IJ}$-field equation (duality equation) in [2], up to an unessential overall factor. Note that all the gravitino-dependent terms are absorbed into the supercovariantized field strengths $\widehat{G}_{\mu\nu}{}^{IJ}$ and $\widehat{\mathcal{P}}_{\mu A}$. This also indicates the internal consistency of our theory. As in [2], eq. (2.11) implies that the vector field $B_\mu{}^{IJ}$ is defined as non-local and non-linear functions of the 248 scalar coordinates for $E_{8(+8)}$. Due to this duality, the vector field $B_\mu{}^{IJ}$ is as physical and propagating as the coset coordinates of $E_{8(+8)}/SO(16)$. Moreover, as in [2], if we multiply both sides of (2.11) by $\epsilon_{\tau\mu\lambda}\partial^\lambda$, making use of the integrability condition (2.9), we get

$$\partial_\nu G^{\mu\nu\, IJ} = \tfrac{1}{4}e^{-1}\epsilon^{\mu\nu\rho}\mathcal{Q}_{\nu\rho}{}^{IJ} + \text{(fermionic terms)} \ . \tag{2.12}$$

This implies that our extra vector field $B_\mu{}^{IJ}$ is physical and propagating in our system. Since this propagating field $B_\mu{}^{IJ}$ has couplings to the $\sigma$-fields as source terms as in (2.12), our gauged $N = 16$ supergravity is equally important as that in [2] with non-trivial difference.

Fourth, the closure on $B_\mu{}^{IJ}$ at the linear order needs a special care. This is because when we apply $[\delta_Q(\epsilon_1), \delta_Q(\epsilon_2)]$ on $B_\mu{}^{IJ}$, we also need the on-shell duality (2.11) leading to $\xi^\nu G_{\nu\mu}{}^{IJ}$ with $\xi^\mu \equiv i(\overline{\epsilon}_2\gamma^\mu\epsilon_1)$ at the linear order. In this process, all the $g$-linear terms in $\delta_Q\psi_\mu$ cancel themselves due to (2.11), and do not contribute to this order. Additionally, a by-product term like $\zeta^{AB}\mathcal{P}_{\mu A}\mathcal{V}_B{}^{IJ}$ in the closure can be regarded as a gauge transformation at the linear order. This closure on $B_\mu{}^{IJ}$ provides another non-trivial consistency check on our total system.

Fifth, there are three relatively independent parameters $g$, $m$ and $\widetilde{m}$ in our theory. Due to the homotopy mapping $\pi_3(SO(16)) = \mathbb{Z}$, $m$ and $\widetilde{m}$ should be quantized as[6]

$$m = \frac{n}{16\pi} \ , \quad \widetilde{m} = \frac{\widetilde{n}}{16\pi} \quad (n, \ \widetilde{n} \in \mathbb{Z}) \ . \tag{2.13}$$

---

[5] Due to the $B_\mu$-field equation $F_{\mu\nu}{}^{IJ} \doteq 0$, there arises no $F_{\mu\nu}$-dependent terms in here.

[6] The factor $1/2$ in front of the Chern-Simons term $mFA + \cdots$ in the lagrangian is the normalization for $SO(16)$ generators, so that we have $16\pi$ in the denominators in (2.13).

Sixth, we can in principle gauge the entire $E_{8(+8)}$ group in $E_{8(+8)}/SO(16)$ by $A_\mu$. However, since the non-compact gauge groups do not have non-trivial $\pi_3$-homotopy mapping, we have gauged only the maximal compact subgroup $SO(16)$ of $E_{8(+8)}$. This is because if the $\pi_3$-homotopy mapping is trivial, the vanishing field strength $F_{\mu\nu}{}^{IJ}$ gives only topologically trivial configurations. Needless to say, we can also gauge any compact subgroup of $SO(16)$ itself, whose $\pi_3$-homotopy mapping is non-trivial.

Seventh, even though the field strength $F_{\mu\nu}{}^{IJ}$ in the gravitino transformation rule in (2.2b) is *not* supercovariantized, this does not pose any problem. Because the difference from the covariantized one is always proportional to the $\lambda$- field which is vanishing by its field equation, so that the on-shell closure of supersymmetry will not be affected. On the other hand, the non-supercovariant field strength is easy to handle for the action invariance confirmation.

Eighth, compared with a model presented in [15], the similarity is that the $SO(N)$ gauge field $A_\mu{}^{IJ}$ is minimally coupled to the gravitino, while its field strength is vanishing by the $BF$-term. The difference is that the present system is more non-trivial, because of the $\sigma$-model physical fields on $E_{8(+8)}/SO(16)$ in addition to other non-propagating fields.

## 3. Application to $SO(9)$-Gauging for $N = 9$ Supergravity

We can rather easily see that a similar mechanism works just fine for other extended supergravity theories with lower $N < 16$. In this paper, we give the example of $N = 9$ supergravity coupled to the $F_{4(-20)}/SO(9)$ $\sigma$-model with an $SO(9)$-gauging. This $N = 9$ supergravity theory is relatively unique, in the sense that it has a simple irreducible structure with the odd number of supersymmetries, with very few analogous examples in any other dimensions. Note also that $N = 9$ supergravity in 3D corresponds to $N > 4$ supergravity in 4D upon simple dimensional reduction [3]. Since only local supersymmetry can exist consistently for $N > 4$ in 4D [18], $N = 9$ supersymmetry in 3D is to be intrinsically *local*. In other words, $N = 9$ supergravity is the simplest example of intrinsically local supersymmetry in 3D.

Corresponding to the case of $E_{8(+8)}/SO(16)$, our $F_{4(-20)}/SO(9)$ has the generators $X^{IJ}$ ($I, J, \cdots = 1, 2, \cdots, 9$) of $SO(9)$, and the coset generators $Y_A$ ($A, B, \cdots = 1, 2, \cdots, 16$), satisfying

$$[X^{IJ}, X^{KL}] = 2\delta^{I[K} X^{L]J} - 2\delta^{J[K} X^{L]I} ,$$
$$[X^{IJ}, Y_A] = -\tfrac{1}{2}(\Gamma^{IJ})_{AB} Y_B , \qquad [Y_A, Y_B] = \tfrac{1}{4}(\Gamma^{IJ})_{AB} X^{IJ} , \tag{3.1}$$

which is just parallel to the $SO(16)$ case [2], except that we need only *undotted* spinorial indices $A, B, \cdots$.

Our field content is $(e_\mu{}^m, \psi_\mu{}^I, \chi_A, \varphi_A, A_\mu{}^{IJ}, B_\mu{}^{IJ}, C_\mu{}^{IJ}, \lambda^I)$. Here we have the indices $A, B, \cdots = 1, 2, \cdots, 16$ for the **16**-spinorial representation, while $I, J, \cdots = 1, 2, \cdots, 9$ for the **9**-vectorial representation both of $SO(9)$. Due to the different chiral spinor structure for $SO(9)$ compared with $SO(16)$, the $\sigma$-model fermion $\chi_A$ has non-dotted index. Since the gauging mechanism is parallel to the $N = 16$ case, we show the total results here: Our lagrangian is[7]

$$\begin{aligned}
e^{-1}\mathcal{L}_9 = &-\tfrac{1}{4}R + \tfrac{1}{2}e^{-1}\epsilon^{\mu\nu\rho}(\overline{\psi}_\mu{}^I \mathcal{D}_\nu(\omega, \mathcal{Q}, A)\psi_\rho{}^I) + \tfrac{1}{4}g^{\mu\nu}\mathcal{P}_{\mu A}\mathcal{P}_{\nu A} \\
&-\tfrac{i}{2}(\overline{\chi}_A \gamma^\mu \mathcal{D}_\mu(\omega, \mathcal{Q}, A)\chi_A) - \tfrac{1}{2}(\overline{\psi}_\mu{}^I \Gamma^I \gamma^\nu \gamma^\mu \chi)_A \mathcal{P}_{\nu A} \\
&+\tfrac{1}{2}ge^{-1}\epsilon^{\mu\nu\rho}B_\mu{}^{IJ}F_{\nu\rho}{}^{IJ} + \tfrac{1}{2}m e^{-1}\epsilon^{\mu\nu\rho}(F_{\mu\nu}{}^{IJ}A_\rho{}^{IJ} + \tfrac{2}{3}g A_\mu{}^{IJ}A_\nu{}^{JK}A_\rho{}^{KI}) \\
&+\tfrac{1}{2}\widetilde{m} e^{-1}\epsilon^{\mu\nu\rho}(H_{\mu\nu}{}^{IJ}C_\rho{}^{IJ} + \tfrac{2}{3}g C_\mu{}^{IJ}C_\nu{}^{JK}C_\rho{}^{KI}) + \tfrac{1}{2}\widetilde{m}(\overline{\lambda}^I \lambda^I) \\
&-\tfrac{1}{8}(\overline{\psi}_\rho{}^I \gamma^{\rho\sigma\tau}\psi_\sigma{}^J)(\overline{\chi}\gamma_\tau \Gamma^{IJ}\chi) + \tfrac{1}{8}(\overline{\psi}_\mu{}^I \gamma_\nu \psi^{\mu J})(\overline{\chi}\gamma^\nu \Gamma^{IJ}\chi) \\
&-\tfrac{1}{8}(\overline{\psi}_\mu{}^I \gamma^\nu \gamma^\mu \psi_\nu{}^J)(\overline{\chi}\chi) + \tfrac{1}{16}(\overline{\chi}\chi)^2 - \tfrac{1}{96}(\overline{\chi}\Gamma^{IJ}\gamma_\mu \chi)^2 \ ,
\end{aligned} \quad (3.2)$$

whose action $I_9 \equiv \int d^3x\, \mathcal{L}_9$ is invariant under supersymmetry

$$\delta_Q e_\mu{}^m = +i(\overline{\epsilon}^I \gamma^m \psi_\mu{}^I) \ , \quad (3.3\text{a})$$

$$\begin{aligned}
\delta_Q \psi_\mu{}^I = &+\mathcal{D}_\mu(\omega, \mathcal{Q}, A)\epsilon^I - \tfrac{i}{4}(\gamma^\nu \epsilon^J)(\overline{\chi}\Gamma^{IJ}\gamma_{\mu\nu}\chi) - \Sigma^{IJ}\psi_\mu{}^J \\
&+ ge^{-1}\epsilon_\mu{}^{\rho\sigma}\widehat{G}_{\rho\sigma}{}^{IJ} + ig\gamma^\nu \epsilon^J \widehat{G}_{\mu\nu}{}^{IJ} \\
&+ \tfrac{i}{2}ge^{-1}\epsilon_\mu{}^{\rho\sigma}\gamma_\rho \epsilon^J \widehat{\mathcal{P}}_{\sigma A}\mathcal{V}_A{}^{IJ} + g\epsilon^J \widehat{\mathcal{P}}_{\mu A}\mathcal{V}_A{}^{IJ} \\
&+ \tfrac{1}{2}ge^{-1}\epsilon_\mu{}^{\rho\sigma}\gamma_\rho \epsilon^J (\overline{\chi}\gamma_\sigma \Gamma^{IJ}\chi) - \tfrac{i}{4}g\epsilon^J(\overline{\chi}\Gamma^{IJ}\gamma_\mu \chi) \ ,
\end{aligned} \quad (3.3\text{b})$$

$$\delta_Q \chi_A = +\tfrac{i}{2}(\Gamma^I)_{AB}(\gamma^\mu \epsilon^I)\widehat{\mathcal{P}}_{\mu B} - \tfrac{1}{4}(\Gamma^{IJ})_{AB}\chi_B \Sigma^{IJ} \ , \quad (3.3\text{c})$$

$$\delta_Q \varphi = +\frac{\Phi}{\sinh \Phi}(\overline{\epsilon}^I \Gamma^I \chi) \equiv \frac{\Phi}{\sinh \Phi} S \ , \quad (3.3\text{d})$$

$$\delta_Q A_\mu{}^{IJ} = +i(\overline{\epsilon}^{[I}\gamma^\nu \widehat{\mathcal{R}}_{\mu\nu}{}^{|J]}) + e^{-1}\epsilon_\mu{}^{\rho\sigma}(\overline{\epsilon}^{[I}\widehat{\mathcal{R}}_{\rho\sigma}{}^{|J]}) + ie^{-1}\epsilon_\mu{}^{\rho\sigma}(\overline{\epsilon}^{[I}\Gamma^{J]}\gamma_\rho \chi)\widehat{\mathcal{P}}_{\sigma A} \ , \quad (3.3\text{e})$$

$$\delta_Q B_\mu{}^{IJ} = +\tfrac{1}{2}(\overline{\epsilon}^K \psi_\mu{}^L)\mathcal{V}^{KL,IJ} - \tfrac{i}{2}(\overline{\epsilon}^K \Gamma^K \gamma_\mu \chi)_A \mathcal{V}_A{}^{IJ} \ , \quad (3.3\text{f})$$

$$\delta_Q C_\mu{}^{IJ} = +i(\overline{\epsilon}^{[I}\gamma_\mu \lambda^{J]}) \ , \quad (3.3\text{g})$$

$$\delta_Q \lambda^I = +\gamma^{\mu\nu}\epsilon^J H_{\mu\nu}{}^{IJ} - \tfrac{i}{2}(\overline{\epsilon}^J \gamma^\mu \psi_\mu{}^J)\lambda^I \ , \quad (3.3\text{h})$$

Since the geometrical structures for the coset $F_{4(-20)}/SO(9)$ are parallel to $E_{8(+8)}/SO(16)$, we do not repeat other relevant equations here.

When the quartic terms for in (3.1) are compared with the $N = 16$ case, only the term

---

[7] For the property of the $\Gamma$-matrices for $SO(9)$ Clifford algebra, see Appendix A.

$(\overline{\chi}\chi)^2$ has a different coefficient. Note also that we do not have the term $(\overline{\chi}\Gamma^{[4]}\chi)^2$, due to the identities (B.2) similar to the $N=16$ case [1].

One crucial identity related to the cancellation of $\chi^3 P$-term in $\delta_Q \mathcal{L}$ is

$$(\overline{\xi}\Gamma^{IJ}\gamma_{\mu\nu}\chi)(\overline{\chi}\Gamma^{IJ}\gamma^{\nu}\chi) \equiv -6(\overline{\xi}\gamma_\mu\chi)(\overline{\chi}\chi) + 2(\overline{\xi}\Gamma^{IJ}\chi)(\overline{\chi}\Gamma^{IJ}\gamma_\mu\chi) \tag{3.4}$$

with $\overline{\xi} \equiv \overline{\epsilon}^K \Gamma^K$, which can be confirmed by the Fierz identity (B.2) in Appendix B.

As in the case of $E_{8(+8)}/SO(16)$ in (2.13), there is non-trivial $\pi_3$-cohomology $\pi_3(SO(9)) = \mathbb{Z}$, so that we have the quantizations

$$m = \frac{n}{16\pi} \quad , \quad \widetilde{m} = \frac{\widetilde{n}}{16\pi} \quad (n,\, \widetilde{n} \in \mathbb{Z}) \ . \tag{3.5}$$

## 4. Yang-Mills Chern-Simons Coupled to $^\forall N$ Extended Supergravity

As careful readers may have noticed, our vector multiplet $(C_\mu{}^{IJ}, \lambda^I)$ can be coupled to arbitrarily large extended supergravities called $\aleph_0$ supergravity [15], in the absence of $\sigma$-model supermultiplets. As a matter of fact, similar models have been given in [15]. However, the field content for a vector multiplet in [15] has both fields in the same adjoint representation of a given group $G$ like $(A_\mu{}^I, \lambda^I)$. The difference here is that $C_\mu{}^{IJ}$ is in the adjoint representation of $SO(N)$, while $\lambda^I$ is in the vector representation. Since we also want to make $SO(N)$ to be local, this system is intrinsically *locally* supersymmetric. This is because the parameter $\epsilon^I$ of supersymmetry is also in the vectorial representation of $SO(N)$, so that we can not impose the global supersymmetry condition such as $\partial_\mu \epsilon^I = 0$, maintaining also the local $SO(N)$ covariance.

Even though this feature sounds rather trivial at first glance, it provides a new concept. Namely, this 'supermultiplet' $(C_\mu{}^{IJ}, \lambda^I)$ has different 'on-shell' degrees of freedom for bosons and fermions, as $N(N-1)/2$ and $N$ respectively. This is possible due to the special feature of a Chern-Simons lagrangian yielding the field strength to vanish, as well as the property of 3D itself where Chern-Simons theory is possible first of all.

We present here such a system of an extra vector multiplet $(C_\mu{}^{IJ}, \lambda^I)$ coupled to $\aleph_0$ extended supergravity plus $SO(N)$ gauge and vector fields: $(e_\mu{}^m, \psi_\mu{}^I, A_\mu{}^{IJ}, B_\mu{}^{IJ})$, as

$$\begin{aligned} e^{-1}\mathcal{L}_{\aleph_0} =& -\tfrac{1}{4}R + \tfrac{1}{2}e^{-1}\epsilon^{\mu\nu\rho}(\overline{\psi}_\mu{}^I D_\nu(\omega, A)\psi_\rho{}^I) \\ & + \tfrac{1}{2}g e^{-1}\epsilon^{\mu\nu\rho} B_\mu{}^{IJ} F_{\nu\rho}{}^{IJ} + \tfrac{1}{2}m e^{-1}\epsilon^{\mu\nu\rho}(F_{\mu\nu}{}^{IJ} A_\rho{}^{IJ} + \tfrac{2}{3} g A_\mu{}^{IJ} A_\nu{}^{JK} A_\rho{}^{KI}) \\ & + \tfrac{1}{2}\widetilde{m} e^{-1}\epsilon^{\mu\nu\rho}(H_{\mu\nu}{}^{IJ} C_\rho{}^{IJ} + \tfrac{2}{3} g C_\mu{}^{IJ} C_\nu{}^{JK} C_\rho{}^{KI}) + \tfrac{1}{2}\widetilde{m}(\overline{\lambda}^I \lambda^I) \ . \end{aligned} \tag{4.1}$$

Needless to say, there is no composite connection in the covariant derivative, such as in $D_\nu(\omega, A)$ by definition. Note also that we do not need the quartic terms independently of $N$, because all the fermions are now only in the vector representations, with no spinorial index for $SO(N)$. In fact, all the explicit quartic terms in (2.1) vanish when the $\sigma$-model fermion $\chi$ is absent, like the present case.

The corresponding action $I_{\aleph_0} \equiv \int d^3x \, \mathcal{L}_{\aleph_0}$ is invariant under supersymmetry

$$\begin{aligned}
\delta_Q e_\mu{}^m &= +i(\overline{\epsilon}^I \gamma^m \psi_\mu{}^I) ~, \\
\delta_Q \psi_\mu{}^I &= +D_\mu(\omega, A)\epsilon^I + g e^{-1}\epsilon_\mu{}^{\rho\sigma}\widehat{G}_{\rho\sigma}{}^{IJ} + ig\gamma^\nu \epsilon^J \widehat{G}_{\mu\nu}{}^{IJ} \\
&\quad + 2im(\gamma^\nu \epsilon^J)F_{\mu\nu}{}^{IJ} + 2me^{-1}\epsilon_\mu{}^{\nu\rho}\epsilon^J F_{\nu\rho}{}^{IJ} ~, \\
\delta_Q A_\mu{}^{IJ} &= +i(\overline{\epsilon}^{[I|}\gamma^\nu \widehat{\mathcal{R}}_{\mu\nu}{}^{|J]}) + e^{-1}\epsilon_\mu{}^{\rho\sigma}(\overline{\epsilon}^{[I|}\widehat{\mathcal{R}}_{\rho\sigma}{}^{|J]}) ~, \\
\delta_Q B_\mu{}^{IJ} &= +(\overline{\epsilon}^{[I}\psi_\mu{}^{J]}) ~, \\
\delta_Q C_\mu{}^{IJ} &= +i(\overline{\epsilon}^{[I}\gamma_\mu \lambda^{J]}) ~, \\
\delta_Q \lambda^I &= +\gamma^{\mu\nu}\epsilon^J H_{\mu\nu}{}^{IJ} - \tfrac{i}{2}(\overline{\epsilon}^J \gamma^\mu \psi_\mu{}^J)\lambda^I ~.
\end{aligned} \quad (4.2)$$

Even though we have added the $SO(N)$ gauge field $A_\mu{}^{IJ}$ with the coupling constant $g$, or the mass parameter $m$, in order to make the result as general as possible, we can delete them by simply putting $g=0$ and/or $m=0$.

Note that this system can have arbitrarily large number $N$ of supersymmetries called $\aleph_0$ supersymmetries [15]. The important aspect here is that such a system is associated with the recent conjecture that a Chern-Simons theory with a certain level on $S^3$ is equivalent to topological string in 2D [16], much like the correspondence between AdS$_3$ and conformal field theory in 4D. In other words, even though the Chern-Simons theory introduced here has vanishing field strength in the 'bulk' of 3D, it has important physical significance at the 2D boundary, similar to the AdS/CFT correspondence [19][9].

## 5. Concluding Remarks

In this paper, we have presented a topologically non-trivial modification of $N = 16$ supergravity in 3D. We have introduced a minimal coupling of an $SO(16)$ gauge field $A_\mu{}^{IJ}$ to the $E_{8(+8)}/SO(16)$ $\sigma$-model, together with an additional vector field $B_\mu{}^{IJ}$, in a combination of a $BF$-theory and Chern-Simons theory. Even though the field strength of the $A_\mu$-field vanishes on-shell, the action is topologically non-trivial due to the homotopy mapping $\pi_3(SO(16)) = \mathbb{Z}$ and the Chern-Simons term. We have also added an additional

Chern-Simons term of an extra gauge field $C_\mu{}^{IJ}$ with an extra topological effects. As an application, we have presented the similar case of $N=9$ supergravity with the $\sigma$-model coset $F_{4(-20)}/SO(9)$. As another application, we have presented a new $\aleph_0$ Chern-Simons theory coupled to extended supergravity with $\forall N$, which was not presented in [15][14].

There are some similarities as well as differences between our gauged system and that in [2]. One important similarity is that the duality relationship (2.11) is exactly the same as in [2], namely, the vector field $B_\mu{}^{IJ}$ is dual to the scalar field strength $\mathcal{P}_{\mu A}$. Therefore, this $B_\mu{}^{IJ}$-field is physical and propagating, so that our gauged system is as non-trivial as the gauging in [2]. The difference is that our system does not have a cosmological constant or the gravitino mass term, while that in [2] does. Instead of a cosmological constant, our system has a topological Chern-Simons term, which leads to non-trivial vacuum configurations. Another difference is that even though the duality relationship is formally the same, our vector $B_\mu{}^{IJ}$ is not a gauge field of $SO(16)$, but its role is played instead by $A_\mu{}^{IJ}$ as an independent field. Moreover, this physically propagating $B_\mu{}^{IJ}$-field has non-trivial couplings to other physical $\sigma$-model fields in its field equation. Due to this non-trivial difference with respect to physical fields, our $N=16$ gauged supergravity is equally important as that in [2].

We have presented in this paper Chern-Simons terms, in particular, for the gauge field $A_\mu{}^{IJ}$ coupling to the $N=16$ gravitino. The supersymmetric partner $\lambda^I$ of $A_\mu{}^{IJ}$ is in the vectorial representation of $SO(16)$. To our knowledge, this is a new supersymmetric Chern-Simons form that has not been covered in the exhaustive studies in refs. [14][15]. Relevantly, if we switch off the $\sigma$-model part for the coset $E_{8(+8)}/SO(16)$, we can formulate such a supersymmetric Chern-Simons term for an arbitrarily large $N$ with no restriction. This is another by-product of our topological gauging of $N=16$ supergravity in 3D.

The non-trivial feature of the gauge field $A_\mu{}^{IJ}$ with vanishing field strength is very peculiar to 3D, because of the non-trivial Chern-Simons term. It is due to the non-trivial $\pi_3$-homotopy of $SO(16)$ or $SO(9)$ that the newly-added Chern-Simons terms with extended supergravity in 3D make stronger sense. However, paradoxically speaking, our results also indicate the possibility that there are some other extensions of 11D supergravity, when topological effects are taken into account. As a matter of fact, such a trial has been presented since 1980's as extra Chern-Simons terms added to 11D [20][21][22]. However, any modification to 11D supergravity, such as higher-order terms, should be also consistent with local supersymmetry. In fact, there has been such a trial on supermembrane corrections to 11D supergravity [23][24][25].

Even though Yang-Mills Chern-Simons theories in 3D look 'trivial', due to their vanishing field strengths in the 'bulk' of 3D, there are lots of non-trivial quantum behaviors, as well as classical topological features. For example, it has been explicitly confirmed that $N = 1$ supersymmetric Chern-Simons theory is finite to all orders in a non-trivial way [26]. Moreover, it has been found that there are non-trivial finite quantum corrections to the Chern-Simons coefficients [27]. From these developments, the model in this paper may well provide a new, unique and non-trivial link between Chern-Simons theories and $N = 16$ maximally extended supergravity in 3D. Also from this viewpoint, our new Chern-Simons model coupled to $\aleph_0$ extended supergravity will be of importance, considering the possible link between Chern-Simons in 3D with topological string in 2D [16].

In this paper, we have also provided the case of $N = 9$ supergravity with the $\sigma$-model coset $F_{4(-20)}/SO(9)$ with non-trivial Clifford algebras. Some of these algebras are very powerful, when dealing with quartic terms, which will be of extra help in the future studies of non-maximal extended supergravities. The case of $N = 9$ extended supergravity is peculiar for the two reasons: First, $N = 9$ is the smallest $N$ in 3D corresponding to $N \geq 5$ supersymmetry in 4D which is *intrinsically* local. Therefore, $N = 9$ is the simplest system with intrinsic local supersymmetry in 3D. Second, the odd dimensionality of orthogonal group $SO(9)$ has very few analogous examples in other higher dimensions.

The importance of the coset $F_{4(-20)}/SO(9)$ comes also from the recent observation that $SO(9)$ might be playing an important role in M-theory [28]. This is because of an interesting analogy between $E_8 \to SO(16)$ and $F_4 \to SO(9)$, due to the coset coordintates of $E_{8(+8)}/SO(16)$ and $F_{4(-20)}/SO(9)$ in the spinorial representations of $SO(16)$ or $SO(9)$, respectively, while $SO(9)$ plays a crucial role as the little group for 11D supergravity as the low energy limit of M-theory [6][9][28].

Our result in this paper has three major important ingredients to be summarized here. First, it is in 3D or lower dimensions, where the generalizations of maximal supergravity by topological terms make stronger sense, due to the non-trivial $\pi_3$-homotopy. Since such modification of maximal supergravity is difficult in $D \geq 4$, it is worthwhile to study possible effects on maximal supergravity in 3D. Second, to put this first point differently, our formulation provides a system that can be a good working ground on the effect of supergravity on non-Abelian Chern-Simons theory, in particular, with the maximal $N = 16$ supersymmetry. The example of $N = 9$ we presented gives a supplementary non-maximal case. Third, our result strongly indicates certain higher-dimensional origin of our new gauging mechanism. For example, M-theory and dualities [6] have lead us to many different generalizations of higher-dimensional origins of certain mechanism in maximally extended supergravity, such as the Killing vector generalization for 11D massive supergravity [29], generalized dimensional reductions [5], or higher-dimensional supergravity theories [12][13]. It will be interesting to see if this leads to new higher dimensional theories in $D \leq 11$ or even $D \geq 12$ [10][11][12].

## Appendix A: $\Gamma$-Matrix Properties for $SO(9)$ Clifford Algebra

In this appendix, we list up some practically useful $\Gamma$-matrix properties for $SO(9)$ Clifford algebra for $N=9$ supergravity. In this appendix, the indices $I, J, \cdots = 1, 2, \cdots, 9$ are for the **9** of $SO(9)$, while $A, B, \cdots, = 1, 2, \cdots, 16$ are for the **16** of $SO(9)$. The symmetry property of the $\Gamma$-matrices for the Clifford algebra for $SO(9)$ is similar to that for $SO(16)$ except for the dottedness for the latter:

$$(\Gamma^I)_{AB} = +(\Gamma^I)_{BA} \quad , \quad (\Gamma^{IJ})_{AB} = -(\Gamma^{IJ})_{BA} \quad , \quad (\Gamma^{[3]})_{AB} = -(\Gamma^{[3]})_{BA} \quad ,$$
$$(\Gamma^{[n]})_{AB} = +(-1)^{n(n-1)/2}(\Gamma^{[n]})_{BA} \quad (0 \leq n \leq 9) \quad , \tag{A.1}$$

which are confirmed by [30]. Since the charge conjugation matrix can be chosen to be the Kronecker's delta: $C_{AB} = \delta_{AB}$ [30], we do not have to distinguish raising/lowering the indices $A, B, \cdots$. For example, $(\overline{\chi}\Gamma^{IJ}\gamma^\mu\chi) \equiv (\Gamma^{IJ})_{AB}(\overline{\chi}_A\gamma^\mu\chi_B)$.

Typical $\Gamma$-algebras are like

$$\Gamma^J\Gamma^I\Gamma^J = -7\Gamma^I \quad , \quad \Gamma^K\Gamma^{IJ}\Gamma^K = +5\Gamma^{IJ} \quad , \quad \Gamma^L\Gamma^{IJK}\Gamma^L = -3\Gamma^{IJK} \quad , \tag{A.2a}$$

$$\Gamma^I\Gamma^{[4]}\Gamma^I = +\Gamma^{[4]} \quad , \quad \Gamma^I\Gamma^{[5]}\Gamma^I = +\Gamma^{[5]} \quad , \quad \Gamma^I\Gamma^{[n]}\Gamma^I = (-1)^n(9-2n)\Gamma^{[n]} \quad , \tag{A.2b}$$

$$\Gamma^{[m]} = +\frac{1}{n!}\epsilon^{[m][n]}\Gamma^{[n]} \quad (m, n = 0, 1, \cdots, 9) \quad , \tag{A.2c}$$

$$\Gamma^{[2]}\Gamma^I\Gamma^{[2]} = -40\Gamma^I \quad , \quad \Gamma^{[2]'}\Gamma^{[2]}\Gamma^{[2]'} = -16\Gamma^{[2]} \quad , \quad \Gamma^{[2]}\Gamma^{[3]}\Gamma^{[2]} = 0 \quad , \tag{A.2d}$$

$$\Gamma^{[2]}\Gamma^{[4]}\Gamma^{[2]} = +8\Gamma^{[4]} \quad , \quad \Gamma^{[3]}\Gamma^I\Gamma^{[3]} = +168\Gamma^I \quad , \quad \Gamma^{[3]}\Gamma^{[2]}\Gamma^{[3]} = 0 \quad , \tag{A.2e}$$

$$\Gamma^{[3]'}\Gamma^{[3]}\Gamma^{[3]'} = -48\Gamma^{[3]} \quad , \quad \Gamma^{[3]}\Gamma^{[4]}\Gamma^{[3]} = +24\Gamma^{[4]} \quad , \quad \Gamma^{[4]}\Gamma^I\Gamma^{[4]} = +336\Gamma^I \quad , \tag{A.2f}$$

$$\Gamma^{[4]}\Gamma^{[2]}\Gamma^{[4]} = -336\Gamma^{[2]} \quad , \quad \Gamma^{[4]}\Gamma^{[3]}\Gamma^{[4]} = -144\Gamma^{[3]} \quad , \quad \Gamma^{[4]'}\Gamma^{[4]}\Gamma^{[4]'} = +144\Gamma^{[4]} \quad . \tag{A.2g}$$

As has been mentioned in the text, the symbols such as $[3]$ stand for the totally antisymmetric $IJK$-indices, and the repeated pairs, such as the $[4]$'s on $\Gamma^{[4]}\Gamma^{[3]}\Gamma^{[4]}$ should be contracted as dummy indices.

Note also that these results are also valid as the usual $\gamma$-matrix algebra in 9D [31][32], because of their 'formal' equivalence, independent of the metric signature *except for* (A.2c).

## Appendix B: Fierz Identities for $SO(9)$ Majorana Spinors

We list up here important relationships associated with $\Gamma$-matrices for the Clifford algebra of $SO(9)$ Satisfying $\{\Gamma^I, \Gamma^J\} = 2\delta^{IJ}I_{16}$, where $I_{16}$ is $16 \times 16$ unit matrix. The most crucial relationships are the Fierz identities for quartic terms: Following [1], suppose we use the symbols $T_i$ $(i = 0, \cdots, 4)$ for

$$T_0 \equiv (\bar{\xi}_A \chi_A)(\bar{\chi}_B \chi_B) \equiv (\bar{\xi}\chi)(\bar{\chi}\chi) ,$$
$$T_1 \equiv \left(\bar{\xi}_A (\Gamma^I)_{AB} \chi_B\right)\left(\bar{\chi}_C (\Gamma^I)_{CD} \chi_D\right) \equiv (\bar{\xi}\Gamma^I\chi)(\bar{\chi}\Gamma^I\chi) ,$$
$$T_2 \equiv \tfrac{1}{2!}\left(\bar{\xi}_A \gamma_\mu (\Gamma^{[2]})_{AB} \chi_B\right)\left(\bar{\chi}_C \gamma^\mu (\Gamma^{[2]})_{CD} \chi_D\right) \equiv \tfrac{1}{2!}(\bar{\xi}\gamma_\mu \Gamma^{[2]}\chi)(\bar{\chi}\gamma^\mu \Gamma^{[2]}\chi) ,$$
$$T_3 \equiv \tfrac{1}{3!}\left(\bar{\xi}_A \gamma_\mu (\Gamma^{[3]})_{AB} \chi_B\right)\left(\bar{\chi}_C \gamma^\mu (\Gamma^{[3]})_{CD} \chi_D\right) \equiv \tfrac{1}{3!}(\bar{\xi}\gamma_\mu \Gamma^{[3]}\chi)(\bar{\chi}\gamma^\mu \Gamma^{[3]}\chi) ,$$
$$T_4 \equiv \tfrac{1}{4!}\left(\bar{\xi}_A (\Gamma^{[4]})_{AB} \chi_B\right)\left(\bar{\chi}_C (\Gamma^{[4]})_{CD} \chi_D\right) \equiv \tfrac{1}{4!}(\bar{\xi}\Gamma^{[4]}\chi)(\bar{\chi}\Gamma^{[4]}\chi) , \quad (\text{B.1})$$

where $\xi_A$ and $\chi_A$ are arbitrary Majorana spinors with the **16**-index $A$. There are only three Fierz identities among them, namely, there are only two independent quartic combinations among $T_0, \cdots, T_4$:

$$T_1 \equiv -3T_0 + \tfrac{1}{3}T_2 , \quad (\text{B.2a})$$
$$T_3 \equiv +24T_0 - T_2 , \quad (\text{B.2b})$$
$$T_4 \equiv -6T_0 - \tfrac{1}{3}T_2 . \quad (\text{B.2c})$$

This implies that there are only two independent $T$'s out of the five quartic combinations: $T_0, \cdots, T_4$. This statement can be confirmed by a method similar to that in [1], namely, we first Fierz each of $T_i$ into the linear combinations of all the $T_i$'s, getting five relationships. Then we symbolize these relations as

$$\mathcal{T} = \mathcal{M}\mathcal{T} , \quad (\text{B.3})$$

with

$$\mathcal{T} \equiv \begin{pmatrix} T_0 \\ T_1 \\ T_2 \\ T_3 \\ T_4 \end{pmatrix} , \quad \mathcal{M} \equiv \tfrac{1}{32}\begin{pmatrix} -1 & -1 & +1 & +1 & -1 \\ -9 & +7 & +5 & -3 & -1 \\ +108 & +60 & +8 & 0 & -12 \\ +252 & -84 & 0 & +8 & -12 \\ -126 & -14 & -14 & -6 & -6 \end{pmatrix} . \quad (\text{B.4})$$

Consider next the eigenvalue equation:

$$\det(\mathcal{M} - xI_{16}) = -\tfrac{1}{8}(2x+1)^3(x-1)^2 , \quad (\text{B.5})$$

meaning that there are three eigenvectors of $\mathcal{M}$ for the eigenvalue $-1/2$, in addition to other twos for the eigenvalue $+1$. Let $\mathcal{A}^{(i)}$ $(i = 1, 2, 3)$ be such three eigenvectors of $\mathcal{M}^T$:

$$\mathcal{M}^T \mathcal{A}^{(i)} = -\tfrac{1}{2}\mathcal{A}^{(i)} \quad (i = 1, 2, 3) . \quad (\text{B.6})$$

Since $\mathcal{T}^T\mathcal{M}^T = \mathcal{T}^T$, we have

$$\mathcal{T}\cdot\mathcal{A}^{(i)} = \mathcal{T}^T\mathcal{A}^{(i)} = \mathcal{T}^T\mathcal{M}^T\mathcal{A}^{(i)} = -\tfrac{1}{2}\mathcal{T}^T\mathcal{A}^{(i)} = -\tfrac{1}{2}\mathcal{T}\cdot\mathcal{A}^{(i)} \ , \tag{B.7}$$

This implies that the inner product $\mathcal{T}\cdot\mathcal{A}^{(i)}$ is zero:

$$\mathcal{T}\cdot\mathcal{A}^{(i)} \equiv 0 \qquad (i = 1,\ 2,\ 3) \ . \tag{B.8}$$

Therefore, finding all the relationships among the $\mathcal{T}$'s is equivalent to finding the eigenvectors $\mathcal{A}^{(i)}$. Following the usual linear algebra technique, we can find that examples of these independent eigenvectors are

$$\mathcal{A}^{(1)} = \begin{pmatrix} 42 \\ 0 \\ 0 \\ -1 \\ 3 \end{pmatrix} \ , \quad \mathcal{A}^{(2)} = \begin{pmatrix} 0 \\ 14 \\ 0 \\ 3 \\ 5 \end{pmatrix} \ , \quad \mathcal{A}^{(3)} = \begin{pmatrix} 0 \\ 0 \\ 7 \\ 3 \\ 12 \end{pmatrix} \ . \tag{B.9}$$

Accordingly, the equation $\mathcal{T}\cdot\mathcal{A}^{(i)}$ for each $i = 1,\ 2,\ 3$ gives the relationships (B.2).

Fierz identities (B.2) explain the absence of the $(\chi\Gamma^{[4]}\chi)^2$-term in the lagrangian (3.2).

# References

[1] N. Marcus and J.H. Schwarz, Nucl. Phys. **B228** (1983) 145.

[2] H. Nicolai and H. Samtleben, hep-th/0010076, Phys. Rev. Lett. **86** (2001) 1686; hep-th/0103032, JHEP **04** (2001) 022.

[3] J. Scherk and J.H. Schwarz, Nucl. Phys. **B153** (1979) 61.

[4] E. Cremmer, B. Julia and N. Scherk, Phys. Lett. **76B** (1978) 409; E. Cremmer and B. Julia, Nucl. Phys. **B159** (1979) 141.

[5] I.V. Lavrinenko, H. Lü and C.N. Pope, hep-th/9710243, Class. Quant. Grav. **15** (1998) 2239; P. Meessen, T. Ortín, hep-th/9806120, Nucl. Phys. **B541** (1999) 195.

[6] C. Hull and P.K. Townsend, Nucl. Phys. **B438** (1995) 109; E. Witten, Nucl. Phys. **B443** (1995) 85; P.K. Townsend, *'Four Lectures on M-Theory'*, in *'Proceedings of ICTP Summer School on High Energy Physics and Cosmology'*, Trieste (June 1996), hep-th/9612121; *'M-theory from its Superalgebra'*, hep-th/9712004.

[7] T. Banks, W. Fischler, S.H. Shenker and L. Susskind, Phys. Rev. **D55** (1997) 5112.

[8] B. de Wit, M. Lüscher and H. Nicolai, Nucl. Phys. **B305** (1988) 545.

[9] *For reviews of M(atrix)-theory, see, e.g.,* A. Bilal, Fort. für Phys. **47** (1999) 5; T. Banks, *'TASI Lecture Note on Matrix Theory'*, hep-th/9911068; W. Taylor IV, *The M(atrix) Model of M-Theory'*, Lectures for NATO school *'Quantum Geometry'* (Iceland 1999), hep-th/0002016; *and references therein*

[10] C. Vafa, Nucl. Phys. **B469** (1996) 403.

[11] I. Bars, Phys. Rev. **D55** (1997) 2373.

[12] H. Nishino, hep-th/9703214, Phys. Lett. **428B** (1998) 85; hep-th/9706148, Phys. Lett. **437B** (1998) 303.

[13] H. Nishino, hep-th/9807199, Nucl. Phys. **B542** (1999) 217.

[14] H. Nishino and S.J. Gates, Jr., Int. Jour. Mod. Phys. **A8** (1993) 3371.

[15] H. Nishino and S.J. Gates, Jr., hep-th/9606090, Nucl. Phys. **B480** (1996) 573.

[16] C. Vafa, *'Unifying Themes in Topological Field Theories'*, hep-th/0005180; R. Dijkgraaf and C. Vafa, *'Matrix Models, Topological Strings, and Supersymmetric Gauge Theories'*, hep-th/0206255; M. Aganagic, M. Marino and C. Vafa, *'All Loop Topological String Amplitudes from Chern-Simons Theory'*, hep-th/0206164.

[17] P. van Nieuwenhuizen, Phys. Rep. **68C** (1981) 189.

[18] *'Supergravity in Diverse Dimensions'*, Vols. **1** & **2**, A. Salam and E. Sezgin, *eds.*, North-Holland, World Scientific (1989); *and references therein.*

[19] I. Klebanov, Nucl. Phys. **B496** (1997) 231, hep-th/9702076; S. Gubser, I. Klebanov and A.A. Tseytlin, Nucl. Phys. **B499** (1997) 217, hep-th/9703040; S. Gubser and I. Klebanov, Phys. Lett. **B 413** (1997) 41, hep-th/9708005; J. Maldacena, Adv. Theor. Math. Phys. **2** (1998) 231, hep-th/9711200; S. Gubser, I. Klebanov and A. Polyakov, Phys. Lett. **B428**

(1998) 105, hep-th/9802109; E. Witten, Adv. Theor. Math. Phys. **2** (1998) 253, hep-th/9802150.

[20] P. van Nieuwenhuizen, Stony Brook preprint, Print-84-0737 (Aug. 1984).

[21] C. Vafa and E. Witten, hep-th/9505053, Nucl. Phys. **B447** (1995) 261.

[22] M.J. Duff, J.T. Liu and R. Minasian, hep-th/9506126, Nucl. Phys. **B452** (1995) 261.

[23] H. Nishino and S.J. Gates, Jr., hep-th/0101037, Phys. Lett. **B508** (2001) 155.

[24] H. Nishino and S. Rajpoot, hep-th/0103224, Phys. Rev. **D64** (2001) 124016.

[25] M. Cederwall, U. Gran, M. Nielsen and B. Nilsson, hep-th/0007035, JHEP **0010** (2000) 041; *'Generalized 11-Dimensional Supergravity'*, hep-th/0010042.

[26] F.R. Ruiz and P. van Nieuwenhuizen, Nucl. Phys. Proc. Suppl. **56B** (1997) 269.

[27] H.-C. Kao, K. Lee and T. Lee, Phys. Lett. **373B** (1996) 94.

[28] P. Ramond, *'Algebraic Dreams'*, hep-th/0112261.

[29] E. Bergshoeff, Y. Lozano and T. Ortin, hep-th/9712115, Nucl. Phys. **B518** (1998) 363.

[30] T. Kugo and P.K. Townsend, Nucl. Phys. **B211** (1983) 157.

[31] S.J. Gates, Jr., H. Nishino and E. Sezgin, Class. and Quant. Gr. **3** (1986) 21.

[32] H. Nishino and S. Rajpoot, hep-th/0207246, CSULB-PA-02-3.